\documentclass[12pt,a4paper]{article}
\pdfoutput=1
\usepackage{jheppub}
\usepackage{feynmp}
\usepackage{caption}          
\usepackage{subcaption}       
\usepackage{shuffle}          

\def\be{\begin{equation}}
\def\ee{\end{equation}}
\def\bea{\begin{eqnarray}}
\def\eea{\end{eqnarray}}
\def\beal{\begin{equation}\begin{aligned}}
\def\eeal{\end{aligned}\end{equation}}
\def\nn{\nonumber}

\def\bra#1{\langle #1|}

\def\sqket#1{|#1]}
\def\braket#1{\langle #1 \rangle}

\def\spA#1#2{\langle #1 #2 \rangle}
\def\spB#1#2{[#1 #2 ]}

\def\Res_#1{\operatorname*{Res}_{#1}}

\def\d{\mathrm{d}}

\def\cN{\mathcal{N}}
\def\tf{\tilde{f}}

\def\eqn#1{eq.~\eqref{#1}}
\def\eqns#1#2{eqs.~\eqref{#1} and~\eqref{#2}}

\def\fig#1{figure~{\ref{#1}}}

\def\Fig#1{Figure~{\ref{#1}}}

\def\tab#1{table~{\ref{#1}}}

\def\Tab#1{Table~{\ref{#1}}}

\def\sec#1{section~{\ref{#1}}}

\def\Sec#1{Section~{\ref{#1}}}

\def\rcite#1{ref.~\cite{#1}}

\def\usedelta#1{\includegraphics[scale=1.0,trim=0 8 0 0]{graphs/#1.pdf}}
\def\usegraph#1#2{\includegraphics[scale=1.0,trim=0 #1 0 0]{graphs/#2.pdf}}

\def\eps{\epsilon}
\def\SpDenom5{\braket{12}\braket{23}\braket{34}\braket{45}\braket{51}}

\def\fmfsettings{
    \fmfset{thin}{1.2pt}
    \fmfset{arrow_len}{7pt}
    \fmfset{arrow_ang}{20}
    \fmfset{wiggly_len}{7pt}
    \fmfset{curly_len}{6pt}
    \fmfset{dash_len}{8pt}
    \fmfset{dot_len}{3pt}
    \fmfset{zigzag_len}{1mm}
    \fmfset{zigzag_width}{.5thick}
}

\title{A Complete Two-Loop, Five-Gluon Helicity Amplitude in Yang-Mills Theory}

\author[a]{Simon Badger,}
\author[a]{Gustav Mogull,}
\author[a]{Alexander Ochirov,}
\author[a,b]{Donal O'Connell\,}

\affiliation[a]{Higgs Centre for Theoretical Physics, School of Physics and Astronomy,\\ The University of Edinburgh, Edinburgh EH9 3JZ, Scotland, UK}
\affiliation[b]{Kavli Institute for Theoretical Physics,  \\University of California, Santa Barbara, CA 93106-4030, USA}

\emailAdd{sbadger@staffmail.ed.ac.uk}
\emailAdd{g.mogull@ed.ac.uk}
\emailAdd{alexander.ochirov@ed.ac.uk}
\emailAdd{donal@staffmail.ed.ac.uk}

\abstract{
We compute the integrand of the full-colour, two-loop, five-gluon scattering amplitude
in pure Yang-Mills theory with all helicities positive, using generalized unitarity cuts.
Tree-level BCJ relations, satisfied by amplitudes appearing in the cuts, allow us
to deduce all the necessary non-planar information for the full-colour amplitude from known planar data.
We present our result in terms of irreducible numerators,
with colour factors derived from the multi-peripheral colour decomposition.
Finally, the leading soft divergences are checked to reproduce
the expected infrared behaviour.
}

\preprint{Edinburgh 2015/16}

\begin{document}
\maketitle
\flushbottom

\section{Introduction}
\label{sec:introduction}

Precision measurements at the Large Hadron Collider provide
detailed information about the nature of the strong interaction
and its role within the Standard Model.
With new data already arriving from Run II,
there is a growing need for higher-precision theoretical predictions
for a variety of different observables.
Recent years have seen considerable progress in this respect.
Our ability to make predictions in perturbative QCD
now covers most of the relevant $2\to2$ scattering processes
at next-to-next-to-leading order (NNLO) precision 
\cite{%
Catani:2011qz,%
Czakon:2013goa,Czakon:2014xsa,%
Gehrmann:2014fva,Grazzini:2015nwa,Grazzini:2013bna,Cascioli:2014yka,%
Boughezal:2015dva,%
Chen:2014gva,Boughezal:2015aha,Boughezal:2015dra,Glover:2001af,Ridder:2015dxa},
as well as the example of inclusive Higgs production at N${}^3$LO
\cite{Anastasiou:2015vya}. Despite this,
processes with more than two particles in
the final state remain beyond the reach of current NNLO methods.


Next-to-leading order (NLO) corrections to high multiplicity final states are
by now commonplace in phenomenological studies. Such computations are possible
thanks to automated techniques, which make use of integrand reduction
\cite{Ossola:2006us}, recursive techniques \cite{Cascioli:2011va},
(generalized) unitarity cuts
\cite{Bern:1994zx,Bern:1994cg,Britto:2004nc,Giele:2008ve} and the known basis
of scalar integrals.
Processes with up to five coloured partons \cite{Bern:2013gka,Badger:2013yda}
in the final state are feasible using on-shell methods that, by only
working with the physical degrees of freedom, are efficient at controlling the
complexity.


Multi-loop calculations of $2\to2$ scattering processes
\cite{Anastasiou:2000ue,Anastasiou:2000kg,Anastasiou:2001sv,%
Glover:2001af,Garland:2002ak,Garland:2001tf,Gehrmann:2011aa,%
Gehrmann:2015ora,vonManteuffel:2015msa,Caola:2015ila,Caola:2014iua}
have been quite successful using the more traditional approaches of Feynman
diagrams and integration-by-parts identities (IBPs) \cite{Chetyrkin:1981qh,Tkachov:1981wb}, though there are notable
exceptions using unitarity cutting techniques
\cite{Bern:2000dn,Bern:2001df,Bern:2002zk,Bern:2002tk,Bern:2003ck,DeFreitas:2004tk}. The main
focus for these processes has been in the evaluation of the resulting master
integrals. At higher multiplicity, rapid growth in the complexity of the
Feynman diagram representation motivates an alternative approach that makes use
of the lessons learned during the automation of one-loop computations.

Two methods have already been explored in this direction. The first of these is maximal
unitarity \cite{Kosower:2011ty}, which generalises the cutting techniques of
Britto, Cachazo and Feng \cite{Britto:2004nc} and Forde \cite{Forde:2007mi} to
compute the rational coefficients of the master integrals, incorporating
information from IBPs. Maximal unitarity has been used to look at maximal cuts for a variety of high-multiplicity examples in four dimensions \cite{%
CaronHuot:2012ab,Johansson:2012zv,Johansson:2013sda,Sogaard:2013yga,%
Sogaard:2013fpa,Sogaard:2014ila,Sogaard:2014oka,Sogaard:2014jla}.
The second approach extends the integrand reduction program of Ossola,
Papadopoulos and Pittau \cite{Ossola:2006us}. The initial steps in this direction
\cite{Mastrolia:2011pr,Badger:2012dp} have now developed into a deeper understanding using the
language of computational algebraic geometry
\cite{Zhang:2012ce,Mastrolia:2012an,Badger:2012dv,Mastrolia:2012wf,Mastrolia:2013kca}. The $D$-dimensional
extension of this method has also been understood and applied in the context of
the planar two-loop, five-gluon amplitude in QCD with all helicities positive (all plus) \cite{Badger:2013gxa}.

In the context of supersymmetric theories, computational 
methods based on an analysis of unitarity cuts have  
enabled a large number of high-loop computations.
Other methods have also been developed, mainly in the context of these
simplified theories. For example, the colour-kinematics duality of Bern, 
Carrasco and Johansson (BCJ) \cite{Bern:2008qj,Bern:2010ue} has been 
successfully exploited to find the complete colour-dressed four-loop, four-gluon
amplitude in $\cN=4$ supersymmetric-Yang-Mills (sYM) \cite{Bern:2010tq}. The two-loop, five-gluon amplitude,
computed in $\cN=4$ sYM in \rcite{Bern:2006vw}, has since been extended to the non-planar sector
and cast into a complete set of numerators satisfying colour-kinematics duality \cite{Carrasco:2011mn}.
The integrands in planar $\cN=4$ sYM are known to all loop orders \cite{ArkaniHamed:2010kv,Bourjaily:2015jna},
and recent studies indicate that this simplicity may extend to the non-planar sector \cite{Arkani-Hamed:2014bca,Franco:2015rma}.

The observation that $\cN=4$ sYM theory and the all-plus sector of
QCD\footnote{The all-plus sector is equivalent to self-dual Yang-Mills at one loop.} are related
by a dimension-shifting relation \cite{Bern:1996ja} suggests that the all-plus amplitude
at two loops could be a useful testing ground for new techniques. Indeed, at two loops the planar sector of the all-plus amplitude was observed to be related to the $\cN=4$ amplitude at the integrand
level in a similar pattern to the one-loop story, although additional corrections to the $\cN=4$ sector appeared in the form of
one-loop squared (or butterfly) topologies \cite{Badger:2013gxa}.
This fact prompts
the question as to how much the techniques applied in supersymmetric cases
may help to simplify QCD applications. 


In this article, we complete the computation of the two-loop, five-gluon,
all-plus helicity amplitude including the non-planar sector. In order to
deal with the increase in complexity of the full colour amplitude, we introduce
a method to find compact colour decompositions that make full use of the
underlying Kleiss-Kuijf (KK) relations \cite{Kleiss:1988ne} in a similar way to the previous treatment at
tree level and one loop by Del Duca, Dixon and Maltoni (DDM) \cite{DelDuca:1999ha,DelDuca:1999rs}.
We then further exploit the on-shell construction of the irreducible numerators
to show that all of the non-planar cuts can be obtained from the planar cuts. This is reminiscent of the colour-kinematics duality,
and indeed we employ the BCJ relations~\cite{Bern:2008qj} at tree level
to relate planar and non-planar cuts.


Our paper is organised as follows. We first present the colour decomposition
of the all-plus, two-loop amplitude, exploiting the multi-peripheral decomposition of the underlying
tree-level amplitudes. In the next section
we describe how the complete kinematic structure can be constructed using knowledge from the planar sector and tree-level identities. After describing a worked example, we present compact results
for the full integrand. We perform checks of the universal soft behaviour of the amplitude by evaluating the
leading $\mathcal{O}(\eps^{-2})$ poles of the integrals in the dimensional regularisation parameter $\eps$.
Finally, we draw some conclusions and discuss some future directions.

\section{Review of irreducible numerators}
\label{sec:irreducible}

We follow a multi-loop integrand reduction algorithm \cite{Mastrolia:2011pr,Zhang:2012ce,Mastrolia:2012an,Badger:2013gxa,Badger:2013sta}
which uses multivariate polynomial division to find an integrand representation of the two-loop amplitude. This section
is intended as a brief overview of the approach; we encourage the reader to refer to the literature for more detailed information.

An integrand-reduced two-loop amplitude has the form
\begin{align}
   \mathcal{A}_n^{(2)}(\{a_i\},\{p_i\}) =
      i g^{n+2} \int \frac{d^d\ell_1 d^d\ell_2}{(2\pi)^{2d}} \sum_\Gamma
      \frac{\tilde{\Delta}_\Gamma(\{a_i\},\{p_i\}, \ell_1,\ell_2)}
           {\prod_{\alpha\in \Gamma} D_\alpha(\{p_i\},\ell_1,\ell_2)},
  \label{IrredIntegrand}
\end{align}
where the sum over runs over graphs $\Gamma$, which are defined by a specific
set of denominators $D_\alpha$ (the set $\{\alpha\}$ labels the propagators in the graph $\Gamma$.) 
Associated with each graph in the sum is a
colour-dressed irreducible numerator $\tilde{\Delta}_\Gamma$; these are
functions of the external momenta $p_i$, the loop momenta $\ell_1$ and
$\ell_2$, and also of the external colour indices $a_i$. Each
of these numerators has a colour decomposition
\begin{align}
   \tilde{\Delta}_\Gamma(\{a_i\},\{p_i\}, \ell_1,\ell_2) =
     \sum_\sigma \sigma \circ \Big[ C_\Gamma(\{a_i\})\,
         \Delta_\Gamma(\{p_i\}, \ell_1,\ell_2) \Big] ,
\label{eq:irrednumdecomp}
\end{align}
where we must explicitly determine the permutation sum $\sigma$ and
the associated colour factors $C_\Gamma$.  We will present an algorithm to
find a simple colour decomposition for our Yang-Mills amplitudes in the next section.

An irreducible numerator $\Delta_\Gamma(\{p_i\},\ell_1,\ell_2)$ can be written
in terms of monomials of irreducible scalar products (ISPs).  To determine a
set of ISPs, we first choose a spanning set of momenta to expand the scalar
products along the lines of the van Neerven-Vermaseren
basis~\cite{vanNeerven:1983vr}.  By re-expressing the propagators in terms of
this spanning set of scalar products, we can see that many can be written as
linear combinations of propagators and can therefore be removed and pushed down
into simpler topologies. These scalar products are known as reducible scalar products (RSPs.) The remaining scalar products are the ISPs; in general
the propagators will be quadratic functions of them. To find a basis set of ISP
monomials, these additional quadratic relations are removed using polynomial
division with respect to a Gr\"{o}bner basis of the relations. This technique
has been developed in the public code \textsc{BasisDet} \cite{Zhang:2012ce}.
When working in $d=4-2\eps$ dimensions we also include three extra-dimensional
ISPs
\be
   \mu_{ij} = \ell_i^{[-2\eps]}\cdot\ell_j^{[-2\eps]}.
\label{mu}
\ee

Once a basis set of ISP monomials is identified, their rational coefficients
are computed from the generalized unitarity cuts of the amplitude.  As we take
all the propagators contained in a particular graph on shell, the cut amplitude factorises into a product of tree-level amplitudes summed over internal
helicity states. Following our schematic notation we can write this as
\beal
   \text{Cut}_\Gamma & = \bigg[\sum_{h_i} \prod_{\alpha \in \Gamma}
      A^{(0)}(\alpha, \{h_i\}) \bigg]_{\text{cut}(\Gamma)} \\ & =
      \bigg[ \Delta_\Gamma(\{p_i\}, \ell_1, \ell_2) 
    - \sum_{\Gamma'\supset \Gamma}
      \frac{\Delta_{\Gamma'}(\{p_i\}, \ell_1, \ell_2)
            \prod_{\alpha\in \Gamma} D_\alpha(\{p_i\}, \ell_1, \ell_2)}
           {\prod_{\alpha\in \Gamma'} D_\alpha(\{p_i\},\ell_1, \ell_2)}
  \bigg]_{\text{cut}(\Gamma)},
  \label{eq:CutDeltaDefinition}
\eeal
where the trees $A^{(0)}$ are those associated with each vertex in the graph
$\Gamma$. 
Making the distinction between the cut
associated with a graph $\Gamma$ and the irreducible numerator associated with
the same graph is crucial for understanding this construction. The irreducible
numerator contains only that information which is required on the cut
associated with $\Gamma$, and which is not captured by irreducible numerators of
graphs $\Gamma'$ that are ``larger'' than $\Gamma$, in the sense that the
propagators contained in $\Gamma'$ are a proper superset of the propagators
contained in $\Gamma$. In other words, by applying the cuts in a top-down
approach we can isolate each topology systematically subtracting the higher-point singularities. 

We will frequently specify the irreducible
numerator associated with a graph $\Gamma$ as $\Delta(\Gamma)$ for clarity; one
should remember that the function $\Delta(\Gamma)$ depends on loop and external
momenta. Furthermore, throughout this paper we will adopt an index notation for the graph labels
which lists the number of propagators in each of the three two-loop branches $\ell_1$,
$\ell_2$ and $\ell_1+\ell_2$. In addition, we add extra labels to distinguish between topologies
of this type. We follow the convention that the right branch is first index, the left branch the second and finally the
central branch in the last entry. For example, the planar pentagon-box reads
$\Delta_{431} = \Delta(\includegraphics[scale=0.5,trim=0 5 0 -5]{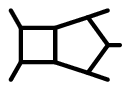})$,
while double box with five legs in a non-planar ordering is
written
$\Delta_{331;5L_2} = \Delta(\includegraphics[scale=0.5,trim=0 5 0 -5]{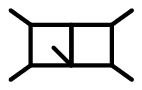})$.
A complete dictionary between this nomenclature
and the graphs relevant for our two-loop, five-point calculation is given in
\Tab{tab:AllGraphs}. 

As shown in \rcite{Badger:2013gxa}, there is only ever a single branch to
the set of solutions to the on-shell equation in $d$ dimensions, which
simplifies the inversion of the system in \eqn{eq:CutDeltaDefinition} to
find the coefficients of the ISP monomials in $\Delta_\Gamma$,
though at the cost of an increased number of monomials
with respect to the four-dimensional case.

There are two important remarks about this construction. The integrand
representation of \eqn{IrredIntegrand} is not unique, and there are
different choices for both the set of ISPs and the set of monomials. Different
sets of spanning vectors will result in different ISPs and the polynomial
division requires a choice of monomial ordering.

In the following sections we will exploit two important consequences of this
approach. Firstly, we will restrict the form of our irreducible numerators to
ensure that the choice of ISPs and monomials satisfy the basic symmetries
required by our colour decompositions. Secondly, we will make use of the
factorisation of irreducible numerators into ordered tree-level amplitudes. These tree
amplitudes satisfy a number of relationships among different orderings. This fact
will allow us to determine all non-planar cuts of the two-loop
all-plus amplitude from the planar irreducible numerators computed in \rcite{Badger:2013gxa}.

\section{Colour decomposition}
\label{sec:colour}

The main result of this work is the construction of the complete five-point,
two-loop, all-plus amplitude in Yang-Mills theory. As
we mentioned in Section~\ref{sec:irreducible}, 
it is necessary to choose a particular colour
decomposition. This decomposition picks a set of colour tensors describing the colour
structure of the amplitude. At the same time, it specifies an associated set of cut
diagrams which must be computed. Each of these cut diagrams is, in turn,
associated with a unique irreducible numerator. Thus the colour decomposition that we pick is of central importance, because it determines the set of irreducible numerators that we need to calculate.

In this section, we describe the general algorithm that we used for constructing an
appropriate colour decomposition of the amplitude, before applying this algorithm to the specific case of the two-loop
five-point amplitude.

\subsection{Multi-peripheral colour decomposition}
\label{sec:reducedexpansion}

\begin{figure}[t]
  \centering
  \includegraphics[width=0.95\textwidth]{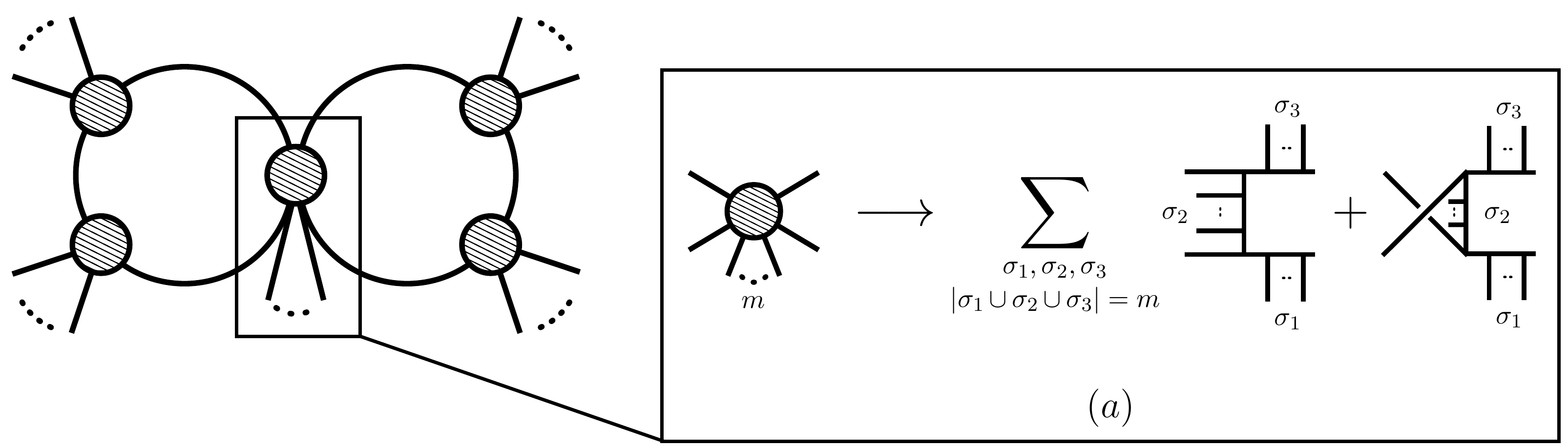}
  \caption{\small Inserting the DDM tree basis into a colour dressed cut of a butterfly topology at two-loops. There are four loop propagators
  in this case, and the insert~$(a)$ shows the result of inserting the DDM tree decomposition fixing the two right legs. The sums run over the permutations of the external legs in the tree-level amplitude.}
  \label{fig:coltopo2}
\end{figure}

Our algorithm is applicable to the general case of an $L$-loop Yang-Mills
amplitude. Following the generalized unitarity principle, we begin by writing
the amplitude as a sum over all colour-dressed cuts. Diagrammatically, these
cuts consist of vertices formed from colour-dressed tree amplitudes which are
joined by on-shell propagators. At two loops the set of colour-dressed cuts
can be classified by two basic topologies: the genuine two-loop topologies are shown in \fig{fig:coltopo1},
and the one-loop squared (or butterfly) topologies in \fig{fig:coltopo2}.

The central idea is to build the loop-level colour decomposition using 
knowledge of the underlying tree-level amplitudes. There are a variety of
well-known presentations of these trees.  We find it convenient to use the DDM
form~\cite{DelDuca:1999ha,DelDuca:1999rs} for the tree amplitudes:
\begin{align}
   {\cal A}^{(0)}_n = -i g^{n-2}
      \sum_{\sigma \in S_{n-2}}
      \tf^{\,a_1 a_{\sigma(2)} b_1} \tf^{\,b_1 a_{\sigma(3)} b_2} \dots
      \tf^{\,b_{n-4} a_{\sigma(n-2)} b_{n-3}}
      \tf^{\,b_{n-3} a_{\sigma(n-1)} a_n} & \nonumber\\ \times
      A^{(0)} (1,\sigma(2),\dots,\sigma(n-1),n) &,
\label{DDM}
\end{align}
where $\tf^{\,abc} = i\sqrt{2}f^{abc}$ are proportional to the standard structure constants in $SU(N_c)$, the gauge group of our Yang-Mills theory.
The main advantage of this form of the amplitude is that
it contains $(n-2)!$ colour structures,
as compared to $(n-1)!$ in the standard trace-based decomposition, for example.
This fact helps to reduce the number of generated diagrams; in particular,
an algorithm based directly on the trace decomposition of tree amplitudes
generates a larger set of diagrams, some of which are rather obscure.

\begin{figure}[t]
\centering
\includegraphics{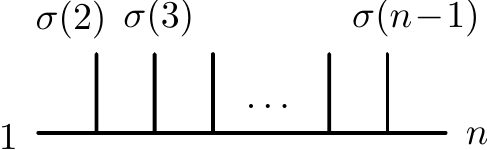}
\caption{\small Multi-peripheral diagram for the colour factors
         in \eqn{DDM}.}
\label{fig:MultiPeripheral}
\end{figure}

Each of the colour structures in the tree decomposition is a string of group
theoretic structure constants $\tf^{abc}$. For an $n$-gluon amplitude
the decomposition is constructed by fixing the position of any two gluons
at either end of this string. The $(n-2)!$ permutations of the remaining gluons
between the ends of the string form the set of colour factors each of which is associated with a colour-ordered tree of the same ordering.
Pictorially, the colour structures look like combs,
as shown in \fig{fig:MultiPeripheral}.

It is straightforward to build the loop colour structure from these DDM tree
colour structures. The loop structure follows directly from the cut diagram:
one simply inserts the DDM trees at the vertices; propagator lines connecting
trees indicate that the ends of the DDM combs at either end of the propagator
have the same colour index to be summed over. Notice, however, that we
must pick two special lines in the DDM form of the tree amplitudes
(corresponding to lines $1$ and $n$ in \fig{fig:MultiPeripheral}.)
These lines are on opposite ends of the DDM colour strings, so one can
informally think of this choice as picking two lines and ``stretching'' the
colour ordered tree between these two ends. We make canonical choices of which
legs to pick as special, depending on the number of propagators that connect
to the three-point amplitude. These choices are:

\begin{figure}[t]
  \centering
  \includegraphics[width=0.8\textwidth]{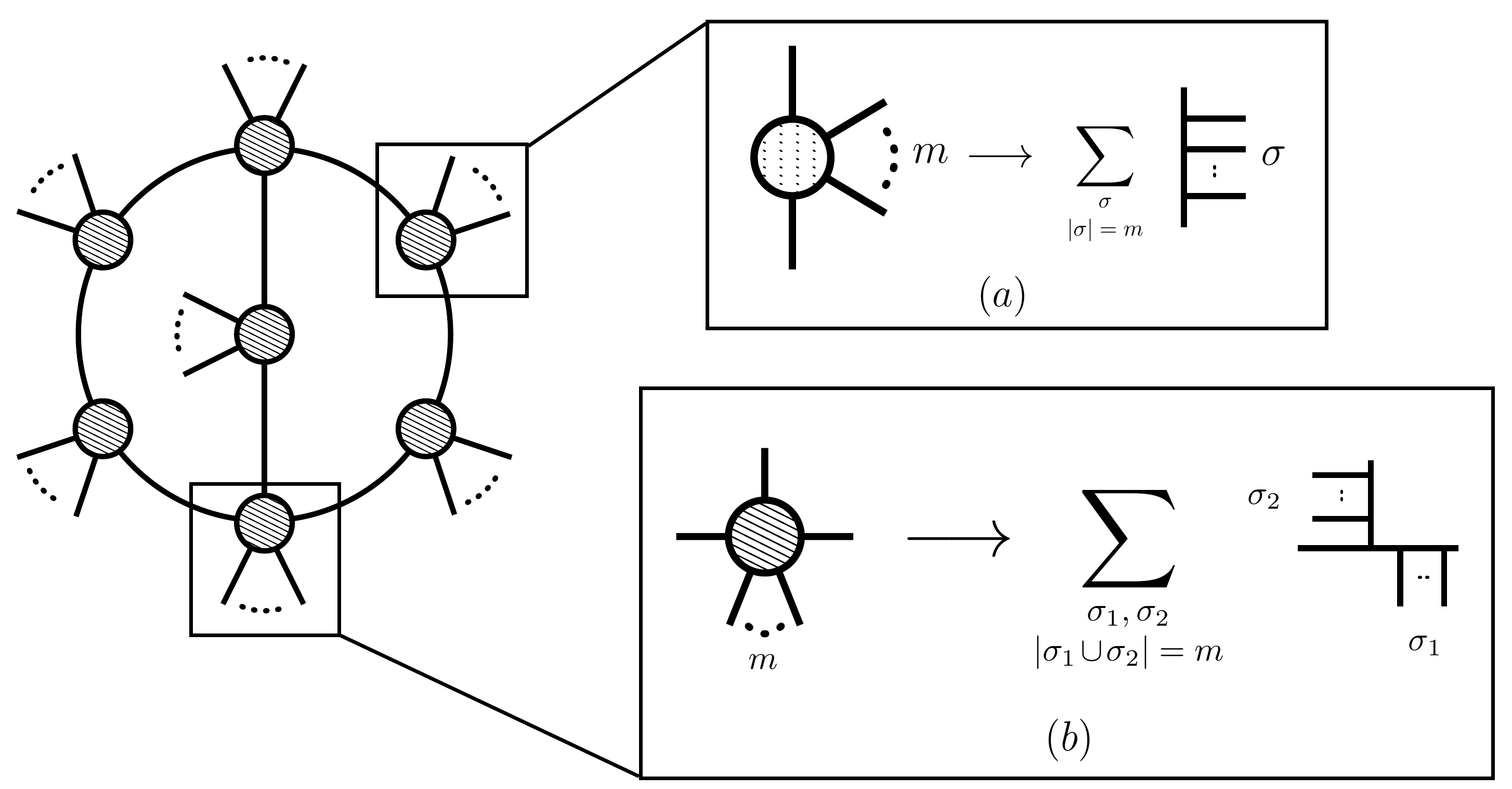}
  \caption{\small Inserting the DDM tree basis into the colour dressed cuts of a two-loop amplitude. The upper insert~$(a)$ shows the simple case of two loop
  propagators, while the lower insert~$(b)$ shows the case of three loop propagators. The sums run over the permutations of the external
  legs in the tree-level amplitude.}
  \label{fig:coltopo1}
\end{figure}

\begin{itemize}
\item Two propagators

In this case, it is natural to ``stretch'' using the two propagators as the
special legs. Thus we build the colour structure by pasting a DDM
multi-peripheral colour structure between the two propagators. We must sum over
every ordering of the external legs. Pictorially, the operation is show in
the upper insert $(a)$ of \fig{fig:coltopo1}.

\item Three propagators

In the case of three propagators, we select two out of the three propagators to
be the special lines in the DDM presentation. Notice that this choice hides
some of the full symmetry of the diagram. In constructing the DDM tree, the
propagators we have selected must be at the end of the multi-peripheral
structure; we must sum over the positions of the other legs. The result is a
sum of diagrams, as shown in the lower insert $(b)$ of \fig{fig:coltopo1}.

\item Four propagators

We again choose two propagators to ``stretch'' the cut amplitude into a DDM
tree. At two loops, we only encounter this case in the butterfly topologies.
We choose upper and lower propagators on the right side of the diagram as special; by symmetry, the result is the same as if we chose upper and lower propagators on the left of the diagram.
The insert of \fig{fig:coltopo2} sketches out the procedure.

\end{itemize}

In this way, we build a set of colour structures. The kinematic structure
associated with each colour structure is easily understood. Each time we insert
a particular DDM colour trace, we also pick up a factor of the associated
colour-ordered tree amplitude. Thus, the orientation of the legs in the kinematic
diagram, associated to an irreducible numerator, is the same as in the colour
structure; of course, the ``stretching'' procedure does not produce new
propagators in the irreducible numerator.

One advantage of using the DDM presentation of the amplitude at tree level is
that the KK relations are automatically satisfied. Our procedure recycles this
property to loop level: we automatically generate a set of colour diagrams
that is KK-independent. Along the way, we generate ordered diagrams for the
kinematics. The same procedure works at $L$ loops;
the amplitude is expressed as
\be
   {\cal A}^{(L)}_n = i^{L-1} g^{n+2L-2} \!\!\!\!\!
                  \sum_{\substack{\text{KK-independent} \\
                                  \text{1PI graphs} \; \Gamma_i}}
                  \int \prod_{j=1}^{L} \frac{\d^d \ell_j}{(2\pi)^d}
                  \frac{1}{S_i}
                  \frac{{C}_i \, \Delta_i(\ell)}{D_i(\ell)} ,
\label{Areduced}
\ee
where $S_i$ are the symmetry factors of the graphs, $D_i$ denote the products
of the (inverse scalar) propagators, and the $\Delta_i$ are irreducible
numerators for appropriate colour factors $C_i$ generated through our
algorithm. Now let us see this procedure at work in the context of the
five-point, two-loop amplitude, which is our main focus.

\subsection{Five-point, two-loop amplitude}
\label{sec:5point2loop}

Now we describe the colour structure of the five-point, two-loop amplitude.
We concentrate on the diagrams that do not vanish in the all-plus case
according to \rcite{Badger:2013gxa}
and our calculations in \sec{sec:kinematic}.
A generic five-point two-loop amplitude can be constructed
by a straightforward extension of the present discussion.

Let us first write the amplitude 
and then explain its content in more detail.
We label $\Delta_i$ and their colour factors
by their diagrams directly in the formula:
\beal \!\!\!\!\!\!
   {\cal A}^{(2)}(1,2,3,4,5) =
   \qquad \qquad \qquad \qquad \qquad \qquad \qquad
   \qquad \qquad \qquad \qquad \qquad \qquad \;\;\! &
   \!\!\!\!\!\!\!\!\!\! \\ \!\!\!\!\!\!\!
   ig^7 \sum_{\sigma \in S_5} \sigma \circ
   I\Bigg[\,
      C\bigg(\usegraph{9}{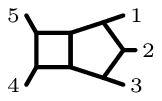}\!\!\!\:\bigg) 
      \Bigg(
      \frac{1}{2} \Delta\bigg(\usegraph{9}{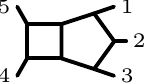}\!\!\!\:\bigg)
 \! + \Delta\bigg(\usegraph{9}{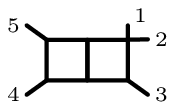}\bigg)
 \! + \frac{1}{2} \Delta\bigg(\usegraph{13}{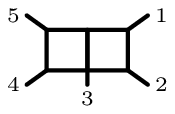}\bigg) &
      \!\!\!\!\!\!\!\!\!\! \\
 \!+\,\frac{1}{2} \Delta\bigg(\usegraph{13}{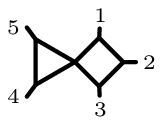}\!\!\!\:\bigg)
 \! + \Delta\bigg(\usegraph{10}{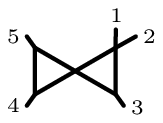}\bigg)
 \! + \frac{1}{2} \Delta\bigg(\usegraph{13}{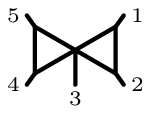}\bigg) &
 \!   \Bigg) \!\!\!\!\!\!\!\!\!\! \\ \!\!\!\!\!
    + C\bigg(\usegraph{9}{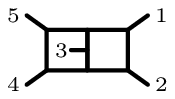}\bigg)
      \Bigg(
      \frac{1}{4} \Delta\bigg(\usegraph{9}{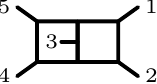}\bigg)
 \! + \frac{1}{2} \Delta\bigg(\usegraph{9}{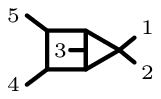}\!\bigg)
 \! + \frac{1}{2} \Delta\bigg(\usegraph{9}{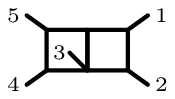}\bigg) & \!\!\!\!\!\!\!\!\!\! \\
 \!-\,\Delta\bigg(\!\!\!\;\usegraph{16}{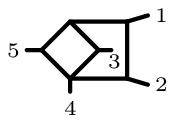}\bigg)
 \! + \frac{1}{4} \Delta\bigg(\usegraph{9}{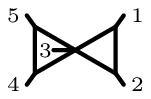}\bigg) &
 \!   \Bigg) \!\!\!\!\!\!\!\!\!\! \\
    + C\bigg(\!\!\!\:\usegraph{9}{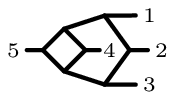}\!\!\!\:\bigg)
      \Bigg(
      \frac{1}{4} \Delta\bigg(\!\!\!\:\usegraph{9}{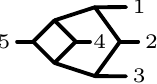}\!\!\!\:\bigg)
 \! + \frac{1}{2} \Delta\bigg(\!\!\!\:\usegraph{8.8}{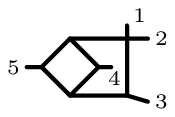}\bigg)
 \!   \Bigg)
 \! + \dots & \,
   \Bigg] , \!\!\!\!\!\!\!\!\!\!
\label{A5point2loop}
\eeal
where the integration operator $I$ acts on every $\Delta_i$ as 
\be
   I[ \Delta_i ] \equiv
      \int \frac{\d^d \ell_1 \d^d \ell_2}{(2\pi)^{2d}} \frac{\Delta_i}{D_i} .
\label{IntMeasure}
\ee
The explicit symmetry factors compensate for the over-counts introduced by the overall sum over permutations of external legs.
For convenience, we recapitulate
all these $\Delta_i$ in \tab{tab:AllGraphs},
where for each irreducible numerator $\Delta_i$ we also show its diagram,
that of its colour factor, as well as the set of its non-equivalent permutations.

\begin{table*}
\centering
\begin{tabular}{lccl}
\hline
\hline
Numerators & Graphs & Colour factors & Permutation sum \\
\hline
$ \Delta_{431} \bigg. $ &
\includegraphics[scale=1.0,trim=0 5 0 -5]{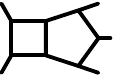} &
\includegraphics[scale=1.0,trim=0 5 0 -5]{graphs/delta431c} &
$S_5/$Vertical flip \\
\hline
$ \Delta_{332} \bigg. $ &
\includegraphics[scale=1.0,trim=0 5 0 -5]{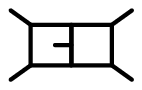} &
\includegraphics[scale=1.0,trim=0 5 0 -5]{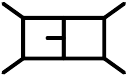} &
$S_5/$Vertical \& horizontal flip \\
\hline
$ \Delta_{422} \bigg. $ &
\includegraphics[scale=1.0,trim=0 5 0 -5]{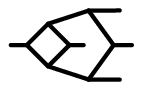} &
\includegraphics[scale=1.0,trim=0 5 0 -5]{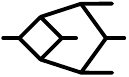} &
$S_5/$Vertical \& diamond flip  \\
\hline
$ \Delta_{331;M_1} \bigg. $ &
\includegraphics[scale=1.0,trim=0 5 0 -5]{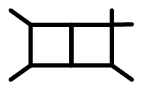} &
\includegraphics[scale=1.0,trim=0 5 0 -5]{graphs/delta431c} &
$S_5$ \\
\hline
$ \Delta_{232;M_1} \bigg. $ &
\includegraphics[scale=1.0,trim=0 5 0 -5]{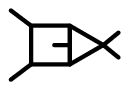} &
\includegraphics[scale=1.0,trim=0 5 0 -5]{graphs/delta332NPc} &
$S_5/$Vertical flip \\
\hline
$ \Delta_{322;M_1} \bigg. $ &
\includegraphics[scale=1.0,trim=0 5 0 -5]{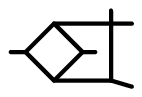} &
\includegraphics[scale=1.0,trim=0 5 0 -5]{graphs/delta422NPc} &
$S_5/$Diamond flip \\
\hline
$ \Delta_{331;5L_1} \bigg. $ &
\includegraphics[scale=1.0,trim=0 5 0 -5]{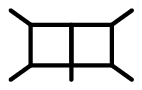} &
\includegraphics[scale=1.0,trim=0 5 0 -5]{graphs/delta431c} &
$S_5/$Horizontal flip \\
\hline
$ \Delta_{331;5L_2} \bigg. $ &
\includegraphics[scale=1.0,trim=0 5 0 -5]{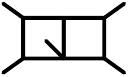} &
\includegraphics[scale=1.0,trim=0 5 0 -5]{graphs/delta332NPc} &
$S_5/$Horizontal flip \\
\hline
$ \Delta_{322;5L_1} \bigg. $ &
\includegraphics[scale=1.0,trim=0 5 0 -5]{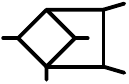} &
\!\!\!\!\!\!$~^-$\includegraphics[scale=1.0,trim=0 5 0 -5]{graphs/delta332NPc} &
$S_5$ \\
\hline
$ \Delta_{430} \bigg. $ &
\includegraphics[scale=1.0,trim=0 5 0 -5]{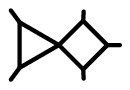} &
\includegraphics[scale=1.0,trim=0 5 0 -5]{graphs/delta431c} &
$S_5/$Vertical flip \\
\hline
$ \Delta_{330;M_1} \bigg. $ &
\includegraphics[scale=1.0,trim=0 5 0 -5]{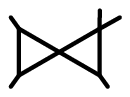} &
\includegraphics[scale=1.0,trim=0 5 0 -5]{graphs/delta431c} &
$S_5$ \\
\hline
$ \Delta_{330;5L_1} \bigg. $ &
\includegraphics[scale=1.0,trim=0 5 0 -5]{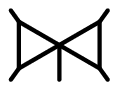} &
\includegraphics[scale=1.0,trim=0 5 0 -5]{graphs/delta431c} &
$S_5/$Horizontal flip \\
\hline
$ \Delta_{330;5L_2} \bigg. $ &
\includegraphics[scale=1.0,trim=0 5 0 -5]{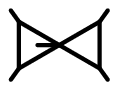} &
\includegraphics[scale=1.0,trim=0 5 0 -5]{graphs/delta332NPc} &
$S_5/$Horizontal \& vertical flip \\
\hline
\hline
\end{tabular}
\caption{\small The irreducible numerators that are nonzero
         for the all-plus five-point two-loop amplitude,
         along with their colour factors and reduced permutation sums.}
\label{tab:AllGraphs}
\end{table*}

The first three graphs in \tab{tab:AllGraphs},
$\Delta_{431}$, $\Delta_{332}$ and $\Delta_{422}$,
are the master diagrams corresponding to the maximal cuts.
They are purely trivalent,
thus their colour factors are unambiguously defined by their proper graphs.

The next three graphs, $\Delta_{331;M_1}$,
$\Delta_{232;M_1}$ and $\Delta_{322;M_1}$,
have a four-point vertex with two external and two internal edges.
The two external legs automatically enter in the permutation sum
with two possible orderings,
hence multi-peripheral subgraphs are naturally obtained
by fixing the internal lines, as in the insert $(a)$ of \fig{fig:coltopo1}.
``Stretching'' the four-point vertex
by these lines gives a master graph for each colour factor.

The following two diagrams,
$\Delta_{331;5L_1}$ and $\Delta_{331;5L_2}$,
share the same graph structure, up to the ordering of the four-point vertex.
The apparent asymmetry introduced by our selecting these two diagrams, 
and omitting the graph with the external leg in the right loop, is an artefact of
our colour decomposition. One could make other choices;
the KK relations satisfied by the trees and their symmetries ensure
that any choice is valid.

To expand the four-point vertex in the ninth graph, $\Delta_{322;5L_1}$,
we fixed the internal lines of the ``diamond'' subdiagram,
hence its colour factor is $C_{332}$,
but with a minus sign due to one flipped vertex.
The other permutation of the four-point vertex
corresponds to the same topology and is present
in the overall permutation sum with the right permuted colour diagram.

The colour factor of the planar graph $\Delta_{430}$ follows in a straightforward manner from our algorithm (see \fig{fig:coltopo2}),
yielding $C_{431}$ as its colour factor.
Its descendant $\Delta_{330;M_1}$ is more interesting,
since it is the only graph in the all-plus case with two four-point vertices.
They can be treated independently by linearity of colour decomposition.
The external one is thus expanded in the same way as in $\Delta_{331;M_1}$,
producing $C_{430}$ as an intermediate step.
Expanding the internal four-point vertex gives $C_{431}$ again.

The last two graphs, $\Delta_{330;5L_1}$ and $\Delta_{330;5L_2}$,
share a five-point vertex. To explain their colour factors,
let us consider the corresponding colour-dressed cut:
\begin{align}
   {\cal C}\text{ut}_{330;5L} = \!\! & \quad\:
      C\bigg(\usegraph{9}{delta431i}\!\!\!\:\bigg)
      \text{Cut}\bigg(\usegraph{13}{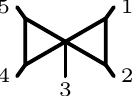}\bigg)
    + C\bigg(\usegraph{9}{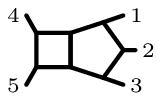}\!\!\!\:\bigg)
      \text{Cut}\bigg(\usegraph{13}{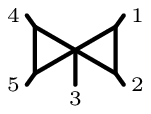}\bigg) \nn \\
  & + C\bigg(\usegraph{9}{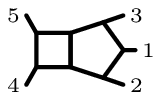}\!\!\!\:\bigg)
      \text{Cut}\bigg(\usegraph{9}{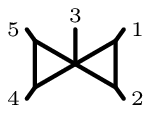}\bigg)
    + C\bigg(\usegraph{9}{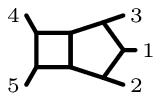}\!\!\!\:\bigg)
      \text{Cut}\bigg(\usegraph{9}{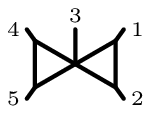}\bigg) \label{Cut3305L1} \\
  & + C\bigg(\usegraph{9}{delta332NPi}\bigg)
      \text{Cut}\bigg(\usegraph{9}{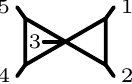}\bigg)
    + C\bigg(\usegraph{9}{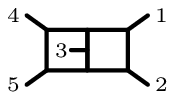}\bigg)
      \text{Cut}\bigg(\usegraph{9}{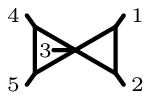}\bigg) . \nn
\end{align} 
We obtain the multi-peripheral decomposition of the five-point vertex
by fixing the two right-hand loop edges
and permuting the other three edges. 
The graphs in the second line can be vertically flipped
to put leg $3$ downstairs to match the presentation in \tab{tab:AllGraphs}.
Obviously, an equivalent decomposition could be achieved
by fixing the loop edges on the left, which would change
the orientation of leg $3$ in the superficially non-planar graphs~$\Delta_{330;5L_2}$.
The $S_5$-summation in \eqn{A5point2loop}
effectively symmetrises the colour structure
over the two choices of multi-peripheral decompositions.

In the present paper we can avoid lower levels of the graph hierarchy
thanks to the simplicity of the fully symmetric helicity configuration,
but it already incorporates all the key elements
of the general colour structure.

\section{Kinematic structure}
\label{sec:kinematic}

With our colour decomposition at hand,
we turn our attention to the kinematic structure of the amplitude.
We need to compute an irreducible numerator associated to each diagram
in \eqn{A5point2loop}; as Frellesvig, Zhang and one of the current authors
have already computed all the planar irreducible numerators~\cite{Badger:2013gxa}, our task is to determine the remaining non-planar numerators. Of course, these numerators can be computed directly from their cuts. However, as we will see, it is easy to determine the complete set of non-planar irreducible numerators for this amplitude from its planar numerators
and the knowledge of tree-level amplitude relations.

\subsection{Non-planar from planar}
\label{sec:npfromp}

The non-planar numerator $\Delta_{332} = \Delta(\includegraphics[scale=0.5,trim=0 5 0 -5]{graphs/delta332NPc})$ can, of course, be obtained directly from its cut.
However, we can avoid calculating  this non-planar cut explicitly by
relating it to a planar cut. We do so in two steps: first, we coalesce two (ordered) three-point trees into a limit
of an ordered four-point tree; then we use the BCJ relations~\cite{Bern:2008qj} satisfied by the ordered four-point tree to reorder the legs until the complete diagram becomes planar.

In more detail, we use the following well-known relation, which is satisfied by on-shell amplitudes in the cuts:
\beal
   A^{(0)}(1,2,-(1\!+\!2)) \, A^{(0)}(1\!+\!2,3,4) \,
  & = \, \big\{ s_{13}\,A^{(0)}(1,3,2,4) \big\} \big|_{s_{12}=0} , \\
      \parbox{70pt}{ \begin{fmffile}{A12A34}
      \fmfframe(10,10)(-10,10){ \fmfsettings \begin{fmfgraph*}(50,25)
            \fmflabel{$1$}{g1}
            \fmflabel{$2$}{g2}
            \fmflabel{$3$}{g3}
            \fmflabel{$4$}{g4}
            \fmfright{g2,g1}
            \fmfleft{g3,g4}
            \fmf{plain}{v12,g1}
            \fmf{plain}{v12,g2}
            \fmf{plain}{g3,v34}
            \fmf{plain}{g4,v34}
            \fmf{plain_arrow,tension=0.50,label.side=left,
                                          label=$1\!+\!2$}{v34,v12}
            \fmfv{decor.shape=circle,decor.filled=full,decor.size=6}{v12,v34}
      \end{fmfgraph*} }
      \end{fmffile} }
  & = \, \left. \!\!\!\: \left\{ s_{13}
      \parbox{50pt}{ \begin{fmffile}{A1324}
      \fmfframe(10,10)(-10,10){ \fmfsettings \begin{fmfgraph*}(30,25)
            \fmflabel{$1$}{g1}
            \fmflabel{$3$}{g2}
            \fmflabel{$2$}{g3}
            \fmflabel{$4$}{g4}
            \fmfright{g2,g1}
            \fmfleft{g3,g4}
            \fmf{plain}{g1,v,g2}
            \fmf{plain}{g3,v,g4}
            \fmfv{decor.shape=circle,decor.filled=full,decor.size=6}{v}
      \end{fmfgraph*} }
      \end{fmffile} }
      \right\} \right|_{s_{12} = 0} ,
\label{BCFW4limit}
\eeal
where  $s_{ij} = (p_i + p_j)^2$ are the standard Mandelstam invariants.
Since this identity is of central importance for us, we present a short proof.
A four-point tree amplitude can be constructed from two three-point amplitudes
using the BCFW recursion relation~\cite{Britto:2004ap,Britto:2005fq}:
\be
   A^{(0)}(1,2,3,4) = \frac{1}{s_{12}} \hat{A}^{(0)}(1,2,-(1\!+\!2))
                                 \hat{A}^{(0)}(1\!+\!2,3,4) ,
\label{BCFW4}
\ee
where hat signs on the right-hand side indicate that the amplitudes
are evaluated on complex kinematics for some BCFW shift of external legs.
The exact complex value of the shifted internal momentum $(\widehat{1\!+\!2})$
is defined by the on-shell condition
\be
   \hat{s}_{12} = s_{12} + \alpha z = 0 ,
\ee
where the precise expression for $\alpha$ depends on the particular BCFW shift.
The key point is that $\hat{s}_{12}$ is a linear function of $z$,
with the property that in the limit $s_{12} \rightarrow 0$,
$z \rightarrow 0$. In this limit \eqn{BCFW4} becomes
\be
   \big\{ s_{12} A^{(0)}(1,2,3,4) \big\} \Big|_{s_{12}=0}
      = A^{(0)}(1,2,-(1\!+\!2)) A^{(0)}(1\!+\!2,3,4) .
\label{BCFW4real}
\ee
Notice that the left-hand side contains a nonzero contribution due to
the pole in $s_{12}$.
Now we can remove the factor of $s_{12}$
on the left-hand side of \eqn{BCFW4real}
by using the BCJ amplitude relation~\cite{Bern:2008qj},
\be
   s_{12} A^{(0)}(1,2,3,4) = s_{13} A^{(0)}(1,3,2,4) .
\label{BCJ4su}
\ee
This proves the identity~\eqref{BCFW4limit}.

We proceed by applying our identity~\eqref{BCFW4limit} to tree amplitudes inside the non-planar cut, rearranging the diagram until it becomes planar.
It is simplest to begin with maximal diagrams, and then to continue to topologies with fewer propagators.
\begin{figure}[t]
\centering
\begin{subfigure}{0.4\textwidth}
\centering
    \parbox{75pt}{  \begin{fmffile}{delta332NP}
    \fmfframe(0,10)(-10,10){ \fmfsettings \begin{fmfgraph*}(70,40) \fmfkeep{delta332NP}
\fmfstraight
\fmfleftn{i}{2}
\fmfrightn{o}{2}
\fmf{plain,tension=3}{i1,v1}
\fmf{plain,tension=3}{i2,v6}
\fmf{plain,tension=3}{v4,o2}
\fmf{plain,tension=3}{v5,o1}
\fmfpoly{plain}{v1,v2,v3,v6}
\fmfpoly{plain}{v2,v5,v4,v3}
\fmffreeze
\fmf{phantom}{v2,v7,v3}
\fmffreeze
\fmf{phantom}{v1,v8,v3}
\fmf{phantom}{v2,v8,v6}
\fmf{plain}{v7,v8}
\fmflabel{$1$}{o2}
\fmflabel{$2$}{o1}
\fmfv{l=$3$,l.a=180,l.d=2}{v8}
\fmflabel{$4$}{i1}
\fmflabel{$5$}{i2}
\fmf{fermion,label=$\ell_1$,label.side=left,tension=0}{v3,v4}
\fmf{fermion,label=$\ell_2$,label.side=right,tension=0}{v3,v6}
    \end{fmfgraph*} }
    \end{fmffile} }
\caption{\small $\Delta_{332}(12345,\ell_1, \ell_2)$ \label{fig:delta332NP}}
\end{subfigure}
\begin{subfigure}{0.4\textwidth}
\centering
    \parbox{75pt}{  \begin{fmffile}{delta422NP}
    \fmfframe(0,10)(-10,10){ \fmfsettings \begin{fmfgraph*}(70,40)
\fmfleft{i1}
\fmfrightn{o}{3}
\fmf{plain,tension=15}{i1,v7}
\fmf{plain,tension=3}{v2,o3}
\fmf{plain,tension=3}{v3,o2}
\fmf{plain,tension=3}{v4,o1}
\fmf{plain,tension=0.5}{v1,v2,v3,v4,v5}
\fmfpoly{phantom}{v5,v4,v3,v2,v1}
\fmfpoly{plain}{v1,v7,v5,v6}
\fmffreeze
\fmf{phantom}{v2,v8}
\fmf{phantom}{v4,v8}
\fmf{plain,tension=0.7}{v6,v8}
\fmflabel{$1$}{o3}
\fmflabel{$2$}{o2}
\fmflabel{$3$}{o1}
\fmfv{l=$4$,l.a=0,l.d=2}{v8}
\fmflabel{$5$}{i1}
\fmf{fermion,label=$\ell_1$,label.side=left,tension=0}{v1,v2}
\fmf{fermion,label=$\ell_2$,label.side=right,tension=0}{v1,v7}
    \end{fmfgraph*} }
    \end{fmffile} }
\caption[b]{\small $\Delta_{422}(12345,\ell_1,\ell_2)$ \label{fig:delta422NP}}
\end{subfigure}
\caption{\small The two non-planar maximal topologies.}
\label{fig:masters}
\end{figure}
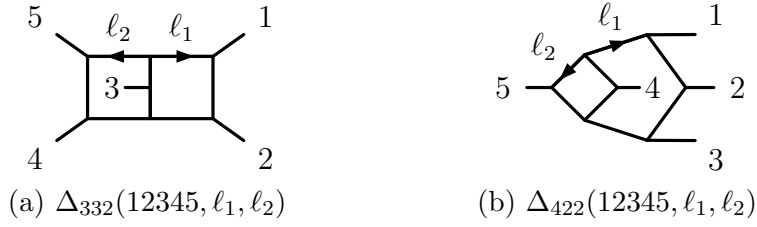
We will work through the calculation of $\Delta_{332}$, displayed in \Fig{fig:masters}, as an example; we computed all non-planar irreducible numerators using the same technique.

The calculation starts at the level of the cuts:
\beal
\!\!\!\!\!\!\!
\text{Cut}_{332} & = ~
\parbox{75pt}{  \begin{fmffile}{cut332NP}
    \fmfframe(0,10)(-10,10){ \fmfsettings \begin{fmfgraph*}(70,40)\fmfkeep{cut332NP}
\fmfstraight
\fmfleftn{i}{2}
\fmfrightn{o}{2}
\fmf{plain,tension=3}{i1,v1}
\fmf{plain,tension=3}{i2,v6}
\fmf{plain,tension=3}{v4,o2}
\fmf{plain,tension=3}{v5,o1}
\fmfpoly{plain}{v1,v2,v3,v6}
\fmfpoly{plain}{v2,v5,v4,v3}
\fmffreeze
\fmf{phantom}{v2,v7,v3}
\fmffreeze
\fmf{phantom}{v1,v8,v3}
\fmf{phantom}{v2,v8,v6}
\fmf{plain}{v7,v8}
\fmfdot{v1,v2,v3,v4,v5,v6,v7}
\fmflabel{$1$}{o2}
\fmflabel{$2$}{o1}
\fmfv{l=$3$,l.a=180,l.d=2}{v8}
\fmflabel{$4$}{i1}
\fmflabel{$5$}{i2}
\fmf{fermion,label=$\ell_1$,label.side=left,tension=0}{v3,v4}
\fmf{fermion,label=$\ell_2$,label.side=right,tension=0}{v3,v6}
    \end{fmfgraph*} }
    \end{fmffile} } 
  = (\ell_1+\ell_2+p_3)^2 ~\left.
\parbox{75pt}{  \begin{fmffile}{cut331M2NP}
    \fmfframe(0,10)(-10,10){ \fmfsettings \begin{fmfgraph*}(70,40)\fmfkeep{cut331M2NP}
\fmfstraight
\fmfleftn{i}{2}
\fmfrightn{o}{2}
\fmf{plain,tension=3}{i1,v1}
\fmf{plain,tension=3}{i2,v6}
\fmf{plain,tension=3}{v4,o2}
\fmf{plain,tension=3}{v5,o1}
\fmfpoly{plain}{v1,v2,v3,v6}
\fmfpoly{plain}{v2,v5,v4,v3}
\fmffreeze
\fmf{phantom}{v1,v7,v3}
\fmf{plain,tension=0.3}{v2,v7}
\fmfdot{v1,v2,v3,v4,v5,v6}
\fmflabel{$1$}{o2}
\fmflabel{$2$}{o1}
\fmfv{l=$3$,l.a=180,l.d=2}{v7}
\fmflabel{$4$}{i1}
\fmflabel{$5$}{i2}
\fmf{fermion,label=$\ell_1$,label.side=left,tension=0}{v3,v4}
\fmf{fermion,label=$\ell_2$,label.side=right,tension=0}{v3,v6}
    \end{fmfgraph*} }
    \end{fmffile} } ~~\right|_{(\ell_1+\ell_2+p_3)^2=0} \\
& = (\ell_1-p_{123})^2 ~\left.
\parbox{75pt}{  \begin{fmffile}{cut3315L}
    \fmfframe(0,10)(-10,10){ \fmfsettings \begin{fmfgraph*}(70,40)\fmfkeep{cut3315L}
\fmfstraight
\fmfleftn{i}{2}
\fmfrightn{o}{2}
\fmfbottomn{b}{1}
\fmf{plain,tension=3}{i1,v1}
\fmf{plain,tension=3}{i2,v6}
\fmf{plain,tension=3}{v4,o2}
\fmf{plain,tension=3}{v5,o1}
\fmfpoly{plain}{v1,v2,v3,v6}
\fmfpoly{plain}{v2,v5,v4,v3}
\fmffreeze
\fmf{plain}{v2,b1}
\fmfdot{v1,v2,v3,v4,v5,v6}
\fmflabel{$1$}{o2}
\fmflabel{$2$}{o1}
\fmflabel{$3$}{b1}
\fmflabel{$4$}{i1}
\fmflabel{$5$}{i2}
\fmf{fermion,label=$\ell_1$,label.side=left,tension=0}{v3,v4}
\fmf{fermion,label=$\ell_2$,label.side=right,tension=0}{v3,v6}
    \end{fmfgraph*} }
    \end{fmffile} } ~~\right|_{(\ell_1+\ell_2+p_3)^2=0} \!\!\!\!\!\!\!\!\!\!
= (\ell_1-p_{123})^2\,\text{Cut}_{331;5L_1}|_{(\ell_1+\ell_2+p_3)^2=0} ,
\!\!\!\!\!\!\!\!\!\!\!\!\!\!\!\!\!
\label{cut332NPdiagram}
\eeal
where $p_{i \dots j} = p_i + \dots + p_j$, and
we understand that all internal helicities are implicitly summed over
while all exposed propagators are cut.
These cuts are decomposed into irreducible numerators as
\begin{subequations}\label{eq:delta332NPBCJ}
\begin{align} & \quad~\:
\text{Cut}_{332} =
\Delta_{332}(12345,\ell_1, \ell_2),
\label{cut332NP} \\ &
\begin{aligned}
\text{Cut}_{331;5L_1}
   = \Delta_{331;5L_1}(12345,\ell_1, \ell_2)
   + \frac{\Delta_{431}(12345,\ell_1, \ell_2)}{(\ell_1-p_{123})^2} & \\
   +  \frac{\Delta_{431}(34512,p_{345}-\ell_2,p_{12}-\ell_1)}
          {(\ell_2-p_{345})^2} & .
\label{eq:cuteqn3315L}
\end{aligned}
\end{align}
\end{subequations}
Using the fact that $(\ell_1-p_{123})^2=-(\ell_2-p_{345})^2$ on this cut,
we see that
\beal
   \Delta_{332}(12345,\ell_1, \ell_2) =
      (\ell_1-p_{123})^2 \Delta_{331;5L_1}(12345,\ell_1,\ell_2)
      +\Delta_{431}(12345,\ell_1, \ell_2) & \\
      -\Delta_{431}(34512,p_{345}-\ell_2,p_{12}-\ell_1) &.
\label{eq:delta332NPos}
\eeal
A similar calculation for $\Delta_{422}$ leads to
\begin{align}
   \Delta_{422}(12345,\ell_1, \ell_2) = \Delta_{431}(12345,\ell_1, \ell_2) .
\label{eq:delta422NPos}
\end{align}

So far the obtained non-planar numerators are valid only on their cuts,
but they can be extended off shell.
We may simply express the numerators in terms of a given set of ISPs and then
define off-shell numerators unambiguously through these ISPs.
The value of a given numerator depends on the choice of ISP basis off shell
(in contrast to the situation on shell, of course).
In this way, we determine a valid set of non-planar irreducible numerators. 
Notice that
the ISP monomial choices made in the planar sector,
such as the higher powers of $\mu_{ij}$
preferred over high powers of $(\ell_i\!\cdot\!p_j)$,
will then be easily translated to the non-planar numerators.

It is very useful
to maintain the symmetries of the underlying graphs in this off-shell continuation.
We achieve this by choosing an appropriate basis of ISPs on a graph-by-graph basis. One engineers the ISP basis such that the loop momentum-dependence of each irreducible numerator is captured by a set of ISPs, which map into one another under the graph symmetries without using any cut conditions. The symmetries of the maximal non-planar graphs, for example, are
\begin{subequations}\label{eq:mastersymmetries}
\begin{align}
   \Delta_{332}(12345,\ell_1,\ell_2) & =
    - \Delta_{332}(21354,p_{12}-\ell_1,p_{45}-\ell_2) = 
      \Delta_{332}(54321,\ell_2,\ell_1) ,
\label{eq:delta332NPsymmetry} \\
   \Delta_{422}(12345,\ell_1,\ell_2) & =
    - \Delta_{422}(32145,p_{123}-\ell_1,p_5-\ell_2) = 
      \Delta_{422}(12354,\ell_1,-\ell_1-\ell_2) .
\label{eq:delta422NPsymmetry}
\end{align}
\end{subequations}

\begin{table*}[t]
\centering
\begin{tabular}{lll} 
\hline
\hline
Graphs & ISPs & RSPs \\
\hline
$\Delta_{332}$ & $\ell_1\cdot(p_5-p_4)$, & $\ell_1^2$, $(\ell_1-p_1)^2$, $(\ell_1-p_{12})^2$, \\
& $\ell_2\cdot(p_1-p_2)$, & $\ell_2^2$, $(\ell_2-p_5)^2$, $(\ell_2-p_{45})^2$, \\
& $(\ell_1-\ell_2)\cdot p_3$ & $(\ell_1+\ell_2+p_3)^2$, $(\ell_1+\ell_2)^2$ \\
\hline
$\Delta_{422}$ & $(\ell_1+2\ell_2)\cdot(p_1-p_3)$, & $\ell_1^2$, $(\ell_1-p_1)^2$, $(\ell_1-p_{12})^2$, $(\ell_1-p_{123})^2$, \\
& $(\ell_1+2\ell_2)\cdot p_2$, & $\ell_2^2$, $(\ell_2-p_5)^2$,\\
& $\ell_1\cdot(p_5-p_4)$ & $(\ell_1+\ell_2+p_4)^2$, $(\ell_1+\ell_2)^2$ \\
\hline
\hline
\end{tabular}
\caption{\small The choices of ISPs and RSPs for the two non-planar masters, 
where the RSPs are chosen as the propagators of the respective graphs.
Additionally, the higher-dimensional ISPs $\mu_{ij}$
are shared by all topologies.}
\label{tab:mastersps}
\end{table*}

These symmetries motivate our choices of ISPs,
given in Table \ref{tab:mastersps}.
For instance, the second symmetry of $\Delta_{332}$ in \eqref{eq:delta332NPsymmetry} leads to a map of ISPs
\begin{align}
  \ell_1\cdot(p_5-p_4) &\leftrightarrow \ell_2\cdot(p_1-p_2) ,\nonumber\\
  (\ell_1-\ell_2)\cdot p_3&\leftrightarrow -(\ell_1-\ell_2)\cdot p_3.
\label{eq:332NPispsym}
\end{align}
After we express loop-momentum dependence
in \eqns{eq:delta332NPos}{eq:delta422NPos}
in terms of the ISPs of \Tab{tab:mastersps},
using the fact that the RSPs (cut propagators) are zero on shell,
we are left with appropriate off-shell irreducible numerators.
These are listed in \Sec{sec:summary}.
Note that the function $\Delta_{422}$ is not the same as the function $\Delta_{431}$ despite the on-shell equation (\ref{eq:delta422NPos}):
different ISPs are chosen to make different off-shell symmetries manifest.

We obtained irreducible numerators for lower-level non-planar diagrams 
in the same way, using the BCJ relations on cuts and extending the results off shell.
For the all-plus amplitude at hand
we find that many lower-level irreducible numerators vanish.
In other words, the higher-level numerators
capture the unitarity cut structure of the full amplitude,
which is given below.

\section{The full-colour five-gluon all-plus amplitude}
\label{sec:summary}

In this section we present a complete summary of all kinematic numerators contributing to the
colour decomposition in \eqn{A5point2loop}. We include both planar \cite{Badger:2013gxa}
and non-planar irreducible numerators computed using the technique described in \Sec{sec:npfromp}.
The result is presented unrenormalized including the dependence on the spin dimension $D_s$ of the gluon,
which is equal to $4$ in the FDH scheme and $4-2\eps$ in CDR \cite{Bern:2002zk}.
The dependence on the extra dimensional ISPs $\mu_{ij} = \ell_i^{[-2\eps]}\cdot\ell^{[-2\eps]}_j$ can be collected into three general
functions,
\begin{subequations} \begin{align}
 & F_1 = (D_s\!-\!2) \big(  \mu_{11} \mu_{22} + (\mu_{11}+\mu_{22})^2
                         +2(\mu_{11}+\mu_{22}) \mu_{12}
                     \big)
       + 16 (\mu_{12}^2 - \mu_{11} \mu_{22}) ,
\label{F1} \\
 & F_2 = 4 (D_s\!-\!2) (\mu_{11}+\mu_{22}) \mu_{12} ,
\label{F2} \\
 & F_3 = (D_s\!-\!2)^2 \mu_{11} \mu_{22} .
\label{F3}
\end{align} \label{Frecap} \end{subequations}
\!\!The remaining coefficients are expressed using the standard spinor-helicity formalism. In particular, we denote
\beal
   \text{tr}_5 & = 4i \epsilon_{\mu\nu\rho\sigma}
                   p_1^\mu p_2^\nu p_3^\rho p_4^\sigma \\
               & = \text{tr}_+(1234)-\text{tr}_-(1234) \\
               & = \spB12\spA23\spA34\spA41 - \spA12\spB23\spA34\spB41 ,
\label{tr5convention}
\eeal
since $\text{tr}_\pm = \text{tr}\left( \tfrac{1}{2}(1\pm\gamma_5)1234\right)$. We also make use of ``spurious"
directions in order to find compact representations of the integrands,
\be
   \omega_{abc}^\mu = \frac{\braket{bc}[ca]}{s_{ab}}
                      \frac{\bra{a}\gamma^\mu\sqket{b}}{2}
                    - \frac{\braket{ac}[cb]}{s_{ab}}
                      \frac{\bra{b}\gamma^\mu\sqket{a}}{2} .
\label{omega}
\ee

The full amplitude reads
\beal \!\!\!\!\!\!\!\!
   {\cal A}^{(2)}(1^+\!,2^+\!,3^+\!,4^+\!,5^+) =
   \qquad \qquad \qquad \qquad \qquad \qquad
   \qquad \qquad \qquad \qquad \qquad \qquad &
   \!\!\!\!\!\!\!\!\!\! \\ \!\!\!\!\!\!\!
   ig^7 \sum_{\sigma \in S_5} \sigma \circ
   I\Bigg[\,
      C\bigg(\usegraph{9}{delta431i}\!\!\!\:\bigg) 
      \Bigg(
      \frac{1}{2} \Delta\bigg(\usegraph{9}{delta431i}\!\!\!\:\bigg)
 \! + \Delta\bigg(\usegraph{9}{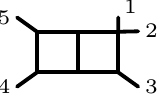}\bigg)
 \! + \frac{1}{2} \Delta\bigg(\usegraph{13}{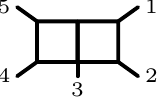}\bigg) &
      \!\!\!\!\!\!\!\!\!\! \\
 \!+\,\frac{1}{2} \Delta\bigg(\usegraph{13}{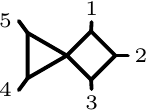}\!\!\!\:\bigg)
 \! + \Delta\bigg(\usegraph{10}{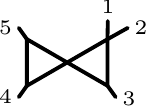}\bigg)
 \! + \frac{1}{2} \Delta\bigg(\usegraph{13}{delta3305Li}\bigg) &
 \!   \Bigg) \!\!\!\!\!\!\!\!\!\! \\ \!\!\!\!\!
    + C\bigg(\usegraph{9}{delta332NPi}\bigg)
      \Bigg(
      \frac{1}{4} \Delta\bigg(\usegraph{9}{delta332NPi}\bigg)
 \! + \frac{1}{2} \Delta\bigg(\usegraph{9}{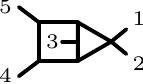}\!\bigg)
 \! + \frac{1}{2} \Delta\bigg(\usegraph{9}{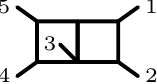}\bigg) & \!\!\!\!\!\!\!\!\!\! \\
 \!-\,\Delta\bigg(\!\!\!\;\usegraph{16}{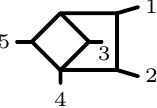}\bigg)
 \! + \frac{1}{4} \Delta\bigg(\usegraph{9}{delta3305L2i}\bigg) &
 \!   \Bigg) \!\!\!\!\!\!\!\!\!\! \\
    + C\bigg(\!\!\!\:\usegraph{9}{delta422NPi}\!\!\!\:\bigg)
      \Bigg(
      \frac{1}{4} \Delta\bigg(\!\!\!\:\usegraph{9}{delta422NPi}\!\!\!\:\bigg)
 \! + \frac{1}{2} \Delta\bigg(\!\!\!\:\usegraph{8.8}{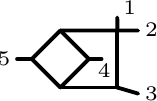}\bigg)
 \!   \Bigg)
   \Bigg] . \!\!\!\!\!\!\!\!\!\!
\label{A5point2loop2}
\eeal
The three maximal graphs are
\beal
   \Delta_{431} & = \Delta\bigg(\usegraph{9}{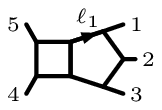}\!\!\!\:\bigg) = 
      -\frac{s_{12}s_{23}s_{45} F_1}
            {\SpDenom5 \text{tr}_5}
            \left(\text{tr}_+(1345) (\ell_1+p_5)^2 + s_{15}s_{34}s_{45} \right),
   \!\!\!\!\! \\
   \Delta_{332} & = \Delta\bigg(\usegraph{9}{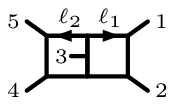}\bigg) =
      \frac{s_{12}s_{45} F_1}
            {4 \SpDenom5 \text{tr}_5}
             \\ & \qquad \qquad \qquad \qquad ~~\: \times
      \Big(\,s_{23} \text{tr}_+(1345) \big(2s_{12} - 4\,\ell_1\!\cdot\!(p_5-p_4)
                                            + 2(\ell_1-\ell_2)\!\cdot\!p_3 \big)
             \\ & \qquad \qquad \qquad \qquad \quad ~~
            -s_{34} \text{tr}_+(1235) \big(2s_{45} - 4\,\ell_2\!\cdot\!(p_1-p_2)
                                            - 2(\ell_1-\ell_2)\!\cdot\!p_3 \big)
             \\ & \qquad \qquad \qquad \qquad \qquad \qquad \qquad \qquad
                      \qquad \qquad \quad ~\;\,
            -4 s_{23}s_{34}s_{15} (\ell_1-\ell_2)\!\cdot\!p_3\,\Big) , \\
   \Delta_{422} & =
   \Delta\bigg(\!\!\!\:\usegraph{9}{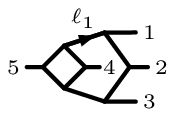}\!\!\!\:\bigg) =
      -\frac{s_{12}s_{23}s_{45} F_1}
            {\SpDenom5 \text{tr}_5}
             \\ & \qquad \qquad \qquad \qquad \,\, \times
      \Big(\,\text{tr}_+(1345) \Big(\ell_1\!\cdot\!(p_5-p_4) - \frac{s_{45}}{2} \Big)
            + s_{15}s_{34}s_{45}\,\Big).
\label{Level0}
\eeal
Meanwhile, the graphs at level 1 are
\beal
   \Delta_{430} & = \Delta\bigg(\usegraph{13}{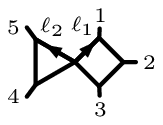}\!\!\!\:\bigg) =
      -\frac{s_{12}\text{tr}_+(1345)}{2 \SpDenom5 s_{13}}
            (2(\ell_1\!\cdot\!\omega_{123}) + s_{23}) \\
      & \qquad \qquad \qquad \qquad \,\, \times
      \left( F_2+F_3 \frac{(\ell_1+\ell_2)^2+s_{45}}
                                   {s_{45}} \right) , \\
   \Delta_{331;5L_1} & = \Delta\bigg(\usegraph{13}{delta3315Li}\bigg) =
      \frac{s_{12}s_{23}s_{34}s_{45}s_{51} F_1}
           {\SpDenom5 \text{tr}_5}, \\ 
   \Delta_{331;5L_2} & = \Delta\bigg(\usegraph{9}{delta3315L2i}\bigg) =
      -\frac{s_{12}s_{45} F_1}{4 \SpDenom5 \text{tr}_5} \\
      & \qquad \qquad \qquad \qquad ~~\: \times
      \big( 2s_{23}s_{34}s_{15} - s_{23}\text{tr}_+(1345)
                                + s_{34}\text{tr}_+(1235) \big), \\
   \Delta_{322;5L_1} & =
   \Delta\bigg(\!\!\!\;\usegraph{16}{delta3225LNPi}\bigg) = 
      -\frac{s_{12} F_1}{2 \SpDenom5 \text{tr}_5} \\
      & \qquad \qquad \qquad \qquad ~~ \times
      \big( s_{23}s_{45}\text{tr}_+(1435)
            -s_{15}s_{34}\text{tr}_+(2453) \big), \\
   \Delta_{331;M_1} & = \Delta\bigg(\usegraph{9}{delta331M1i}\bigg) =
   \Delta_{322;M_1} = \Delta\bigg(\!\!\!\:\usegraph{8.8}{delta322M1NPi}\bigg) =
   \Delta_{232;M_1} = \Delta\bigg(\usegraph{9}{delta232NPi}\!\bigg) \\
      & \qquad \qquad \qquad \quad ~\:\, = 
      -\frac{s_{34}s_{45}^2\text{tr}_+(1235)F_1}{\SpDenom5 \text{tr}_5},
\label{Level1}
\eeal
Finally, the graphs at level 2 are
\begin{align}
   \Delta_{330;M_1} & = \Delta\bigg(\usegraph{10}{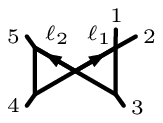}\bigg) =
      -\frac{(s_{45}-s_{12}) \text{tr}_+(1345)}{2 \SpDenom5 s_{13}}
      \left(F_2 + F_3 \frac{(\ell_1+\ell_2)^2 + s_{45}}{s_{45}}
      \right), \nn \\
   \Delta_{330;5L_1} & = \Delta\bigg(\usegraph{13}{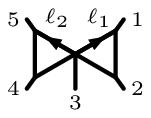}\bigg) =
  -\frac{1}{\SpDenom5} \nn \\ & \!\!\!\!\!\!\!\!\!\!\!\!\!\!\!\!\!\!\!\! \times 
      \Bigg\{ \frac{1}{2}\;\!
             \bigg(\text{tr}_+(1245)
                  -\frac{\text{tr}_+(1345) \text{tr}_+(1235)}{s_{13}s_{35}}
             \bigg) \nn \\ & \!\!\!\!\!\!\!\!\!\times\!
             \bigg(F_2
                  +F_3 \frac{4(\ell_1\!\cdot\!p_3)(\ell_2\!\cdot\!p_3)
                                  +(\ell_1+\ell_2)^2 (s_{12}+s_{45})
                                  + s_{12}s_{45}}
                                 {s_{12}s_{45}}
             \bigg) \nn \\ & \!\!\!\!\!\!\!\!\!\!\!\!
           + F_3 \bigg[\,(\ell_1+\ell_2)^2 s_{15} \\ &\;
   +\text{tr}_+(1235) \bigg( \frac{(\ell_1+\ell_2)^2}{2s_{35}}
                            -\frac{\ell_1\!\cdot\!p_3}{s_{12}}
                      \Big(1 + \frac{2(\ell_2\!\cdot\!\omega_{543})}{s_{35}}
                              + \frac{s_{12}-s_{45}}{s_{35}s_{45}}
                                (\ell_2-p_5)^2
                      \Big)
                      \bigg) \nn \\ &\;
   +\text{tr}_+(1345) \bigg( \frac{(\ell_1+\ell_2)^2}{2s_{13}}
                            -\frac{\ell_2\!\cdot\!p_3}{s_{45}}
                      \Big(1 + \frac{2(\ell_1\!\cdot\!\omega_{123})}{s_{13}}
                              + \frac{s_{45}-s_{12}}{s_{12}s_{13}}
                                (\ell_1-p_1)^2
                      \Big)
                      \bigg)
                      \bigg]
      \Bigg\}, \nn \\
   \Delta_{330;5L_2} & = \Delta\bigg(\usegraph{9}{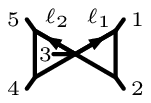}\bigg) =
      \frac{F_3}{2 \SpDenom5 s_{12}} \nn \\ & \quad \times
      \Bigg( (s_{45}-s_{12}) \text{tr}_+(1245)
            -\left( \text{tr}_+(1245)
                   -\frac{\text{tr}_+(1345)\text{tr}_+(1235)}{s_{13}s_{35}}
             \right) 2(\ell_1\!\cdot\!p_3) \nn \\ &
             \qquad \qquad \qquad \qquad \qquad
             -\frac{s_{45} \text{tr}_+(1235)}{s_{35}}
             \left(2(\ell_2\!\cdot\!\omega_{543})
                   +\frac{s_{12}-s_{45}}{s_{45}}(\ell_2-p_5)^2
             \right) \nn \\ &
             \qquad \qquad \qquad \qquad \qquad
             +\frac{s_{12} \text{tr}_+(1345)}{s_{13}}
             \left(2(\ell_1\!\cdot\!\omega_{123})
                   +\frac{s_{45}-s_{12}}{s_{12}}(\ell_1-p_1)^2
             \right)
      \Bigg). \nn
\end{align}

We have found a representation of the full amplitude with no topologies with fewer than six propagators.
We note that there are nonzero cuts at the integrand level,
but the resulting integrals are scaleless and hence zero in dimensional regularisation.
We have checked additional cuts at levels 2 and 3 to ensure that
no nonzero topologies remain.
To find an integrand with this property, the ISPs $(\ell_1\!\cdot\!\omega_{123})$ and $(\ell_2\!\cdot\!\omega_{543})$ in the numerators $\Delta_{330;5L_2}$ and $\Delta_{330;5L_1}$ were upgraded to include terms
proportional to $(\ell_1-p_1)^2$ and $(\ell_2-p_5)^2$.

\section{Checking the soft divergences}

Since the all-plus helicity configuration is zero at tree level,
the universal infrared (IR) structure is the same as that of an ordinary one-loop amplitude. The poles of our two-loop amplitude should therefore
be equivalent to those of the finite one-loop amplitude multiplied by the infrared pole operator
including the sum over colour correlations~\cite{Catani:2000ef}
\begin{align}
  \mathcal{A}^{(2)}&(1^+\!,2^+\!,3^+\!,4^+\!,5^+) =
  \nonumber\\&
  \sum_{\parbox{8mm}{\centering\scriptsize$i,j$\\$j\!\neq\!i$}}
  \frac{c_\Gamma}{\eps^2} \left(\frac{\mu_R^2}{-s_{ij}}\right)^\eps T_i\cdot T_j \circ \mathcal{A}^{(1)}(1^+\!,2^+\!,3^+\!,4^+\!,5^+)
  + \mathcal{O}(\eps^{-1}),
  \label{eq:IRfull}
\end{align}
where we have used the $\circ$ symbol to indicate that the colour matrices (in this purely gluonic case they will all be structure constants) should be inserted into
the colour factors of the one-loop amplitude. The standard loop prefactor is given by \footnote{The normalisation of the integrals in this section
is different by a factor of $i/(4\pi)^{d/2}$ per integration with respect to
the default choices in \textsc{FIESTA} and \textsc{SecDec}.}
\begin{equation}
   c_\Gamma = \frac{\Gamma(1+\eps)\Gamma^2(1-\eps)}
                   {(4\pi)^{2-\eps}\,\Gamma(1-2\eps)}.
\label{eq:cGammaDef}
\end{equation}
The one-loop amplitude to all orders in $\eps$ can be found in ref.~\cite{Bern:1996ja}.

There are a number of difficulties in checking \eqn{eq:IRfull} in full since the five-point planar and non-planar integrals
required are still unknown at this time. Resorting to numerical evaluation, as has been done in the planar case \cite{Badger:2013gxa}, is
computationally prohibitive for two reasons. Firstly, the full colour expansion contains a large number of dimension-shifted
integrals ($\sim\mathcal{O}(1000)$) -- an order of magnitude more than the leading colour terms. Secondly, there is no Euclidean region
for the complete amplitude, and so contour deformation must be performed for many of these integrals, making them more complicated than the
planar cases. This task is probably achievable with public tools like \textsc{FIESTA}~\cite{Smirnov:2013eza} and \textsc{SecDec}~\cite{Borowka:2015mxa}.

There is, however, a much simpler method to check the
leading soft singularities up to $\mathcal{O}(\eps^{-1})$,
which can be done analytically.
In the leading soft limit, the colour correlations drop out of \eqn{eq:IRfull}:
\begin{equation}
  \mathcal{A}^{(2)}(1^+\!,2^+\!,3^+\!,4^+\!,5^+) =
  -\frac{5 N_c c_\Gamma}{\eps^2} \mathcal{A}^{(1)}(1^+\!,2^+\!,3^+\!,4^+\!,5^+)
  + \mathcal{O}(\eps^{-1}).
  \label{eq:IRsoft}
\end{equation}
Clearly this is a weaker check than the full IR poles, but it does require non-trivial properties of the non-planar sector.
The butterfly (one-loop squared) topologies,
are all finite and therefore not relevant for the IR properties.
Scattering amplitudes in the soft or eikonal limit
have many remarkable structures and universal properties.
The interested reader may like to turn to \rcite{White:2015wha}
for a recent introduction to the subject.

\subsection{Evaluating the massless double box in the soft limit}

The extra simplicity in our all-plus loop amplitude
that sets it aside from most two-loop amplitudes
is that the integrals contain at most a single soft divergence,
rather than the maximum double soft divergence.
We can therefore break our loop amplitudes up into sums of regions
with soft singularities and evaluate the amplitude in the limit.
In this limit the integral factorises into a product of two one-loop integrals and can be evaluated to extract the leading $\mathcal{O}(\eps^{-2})$ divergence.

All of the poles of our amplitude are contained in the topologies
proportional to the same dimension shifting numerator:
\begin{equation}
  F_1 = (D_s-2)(3\mu_{11}\mu_{22}+\mu_{11}^2+\mu_{22}^2+2 \mu_{12}(\mu_{11}+\mu_{22})) + 16 (\mu_{12}^2 - \mu_{11}\mu_{22}),
  \label{eq:}
\end{equation}
which also has a simple behaviour,
\begin{equation}
  \underset{\ell_1 \to 0}{\lim} F_1 = (D_s-2)\mu_{22}^2.
\end{equation}

Taking the example of the two-loop double box, we find two soft regions by taking the limit of either loop.
We find a soft singularity whenever we have two adjacent massless legs in one of the loop integrations.
In each case, we factorise into an IR divergent triangle
and a dimension-shifted box:
\begin{subequations} \begin{align}
   I^{4-2\eps}
   \bigg(\,\parbox{37pt}{ \begin{fmffile}{I331l}
      \fmfframe(0,10)(-10,10){ \fmfsettings \tiny
      \begin{fmfgraph*}(35,20)
         \fmfkeep{I331l}
         \fmfleftn{i}{2}
         \fmfrightn{o}{2}
         \fmf{plain,tension=5.5}{i1,v1}
         \fmf{plain,tension=5.5}{i2,v2}
         \fmf{plain,tension=5.5}{o2,v4}
         \fmf{plain,tension=5.5}{o1,v5}
         \fmfpoly{plain}{v6,v3,v2,v1}
         \fmfpoly{plain}{v6,v5,v4,v3}
         \fmf{plain,label=$\ell_1$,label.side=left,
                    label.dist=3pt,tension=0}{v4,v5}
         \fmf{plain,label=$\ell_2$,label.side=right,
                    label.dist=2pt,tension=0}{v2,v1}
         \fmfv{label=1,label.angle=0,label.dist=2pt}{o2}
         \fmfv{label=2,label.angle=0,label.dist=2pt}{o1}
         \fmfv{label=3,label.angle=180,label.dist=2pt}{i1}
         \fmfv{label=4,label.angle=180,label.dist=2pt}{i2}
      \end{fmfgraph*} }
      \end{fmffile} }
   \bigg) [F_1] &
   \xrightarrow{\ell_1 \to 0}
   (D_s-2) I^{4-2\eps}
   \bigg(\;\parbox{28pt}{ \begin{fmffile}{BOX0m}
      \fmfframe(0,10)(-10,10){ \fmfsettings \tiny
      \begin{fmfgraph*}(25,20) \fmfkeep{BOX0m}
         \fmfleftn{i}{2}
         \fmfrightn{o}{2}
         \fmf{plain,tension=2}{i1,v4}
         \fmf{plain,tension=2}{i2,v1}
         \fmf{plain,tension=2}{o1,v2}
         \fmf{plain,tension=2}{o2,v3}
         \fmfpoly{plain}{v1,v4,v2,v3}
         \fmfv{label=1,label.angle=0,label.dist=2pt}{o2}
         \fmfv{label=2,label.angle=0,label.dist=2pt}{o1}
         \fmfv{label=3,label.angle=180,label.dist=2pt}{i1}
         \fmfv{label=4,label.angle=180,label.dist=2pt}{i2}
      \end{fmfgraph*} }
   \end{fmffile} }
  \bigg) [\mu_{22}^2] \,
  I^{4-2\eps}
  \bigg(\!\parbox{28pt}{ \begin{fmffile}{TRI1mL}
      \fmfframe(0,10)(-10,10){ \fmfsettings \tiny
      \begin{fmfgraph*}(25,20) \fmfkeep{TRI1mL}
         \fmfleftn{i}{1}
         \fmfrightn{o}{2}
         \fmf{dbl_plain,tension=2}{i1,v1}
         \fmf{plain,tension=2}{o1,v2}
         \fmf{plain,tension=2}{o2,v3}
         \fmfpoly{plain}{v1,v2,v3}
         \fmfv{label=1,label.angle=0,label.dist=2pt}{o2}
         \fmfv{label=2,label.angle=0,label.dist=2pt}{o1}
      \end{fmfgraph*} }
   \end{fmffile} }
   \bigg) ,
   \\
   I^{4-2\eps}
   \bigg(\,\parbox{37pt}{ \tiny \fmfreuse{I331l} } \bigg) [F_1] &
   \xrightarrow{\ell_2 \to 0}
   (D_s-2) I^{4-2\eps}
   \bigg(\;\parbox{28pt}{ \begin{fmffile}{TRI1mR}
      \fmfframe(0,10)(-10,10){ \fmfsettings \tiny
      \begin{fmfgraph*}(25,20) \fmfkeep{TRI1mR}
         \fmfleftn{i}{2}
         \fmfrightn{o}{1}
         \fmf{dbl_plain,tension=2}{o1,v1}
         \fmf{plain,tension=2}{i1,v2}
         \fmf{plain,tension=2}{i2,v3}
         \fmfpoly{plain}{v1,v3,v2}
         \fmfv{label=3,label.angle=180,label.dist=2pt}{i1}
         \fmfv{label=4,label.angle=180,label.dist=2pt}{i2}
      \end{fmfgraph*} }
   \end{fmffile} }\!\!\!\!\:
  \bigg) \,
  I^{4-2\eps}
  \bigg(\;\parbox{28pt}{ \tiny \fmfreuse{BOX0m} } \bigg) [\mu_{11}^2] .
\end{align} \end{subequations}
Recalling the one-loop integrals,
\begin{align}
  & I^{4-2\eps}
   \bigg(\;\parbox{28pt}{ \tiny \fmfreuse{BOX0m} } \bigg) [\mu_{11}^2] =
   - \frac{i}{(4\pi)^2}\frac{1}{6} + \mathcal{O}(\eps) , \\
    & I^{4-2\eps}
   \bigg(\!\parbox{28pt}{ \tiny \fmfreuse{TRI1mL} } \bigg) =
      \frac{i c_\Gamma}{\eps^2}\left( -s_{12} \right)^{-1-\eps} =
      - \frac{i}{(4\pi)^2}\frac{1}{s_{12}\eps^2} + \mathcal{O}(\eps^{-1}) ,
\end{align}
and summing the two regions, we quickly arrive at the result:
\be
   I^{4-2\eps}
   \bigg(\,\parbox{37pt}{ \begin{fmffile}{I331}
      \fmfframe(0,10)(-10,10){ \fmfsettings \tiny
      \begin{fmfgraph*}(35,20)
         \fmfleftn{i}{2}
         \fmfrightn{o}{2}
         \fmf{plain,tension=5.5}{i1,v1}
         \fmf{plain,tension=5.5}{i2,v2}
         \fmf{plain,tension=5.5}{o2,v4}
         \fmf{plain,tension=5.5}{o1,v5}
         \fmfpoly{plain}{v6,v3,v2,v1}
         \fmfpoly{plain}{v6,v5,v4,v3}
         \fmfv{label=1,label.angle=0,label.dist=2pt}{o2}
         \fmfv{label=2,label.angle=0,label.dist=2pt}{o1}
         \fmfv{label=3,label.angle=180,label.dist=2pt}{i1}
         \fmfv{label=4,label.angle=180,label.dist=2pt}{i2}
      \end{fmfgraph*} }
      \end{fmffile} }
   \bigg) [F_1] =
   -\frac{1}{(4\pi)^4} \frac{D_s-2}{3 s_{12} \eps^2} + \mathcal{O}(\eps^{-1}) .
\ee

\subsection{Soft divergences of the five-point integrals}

By following the method described in the previous section,
we have derived the complete set of integrals
required for the $\mathcal{O}(\eps^{-2})$ part of the amplitude.
All of the integrals have been checked numerically
using the sector decomposition methods implemented in \textsc{FIESTA}~\cite{Smirnov:2013eza} and \textsc{SecDec}~\cite{Borowka:2015mxa}.
Some of these integrals have been computed long ago in $4-2\eps$ dimensions
and can be used to write the full integrals including finite terms
via the dimensional reduction identities implemented in \textsc{LiteRed}~\cite{Lee:2012cn}
and IBP relations from \textsc{FIRE5}~\cite{Smirnov:2014hma}.\footnote{We thank
Claude Duhr for providing his own computation of the integrals
for $e^+e^-\to3j$ \cite{Gehrmann:2000zt,Gehrmann:2001ck}.}
We have performed this task for the planar double box with an off-shell leg.
Thus we arrived at the following soft limits for the integrals:
\begin{subequations} \begin{align}
 & I^{4-2\eps} \bigg(\usegraph{9}{delta431i}\!\!\!\:\bigg) [F_1] =
    \frac{1}{(4\pi)^4} \frac{D_s-2}{3 s_{12}s_{23}\eps^2} + \mathcal{O}(\eps^{-1}) , \\
 & I^{4-2\eps} \bigg(\usegraph{9}{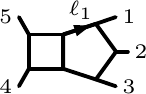}\!\!\!\:\bigg) [F_1\,(\ell_1\!\cdot\!p_5)] =
    \frac{1}{(4\pi)^4} \frac{(D_s-2)(2 s_{15}+s_{25})}{12 s_{12}s_{23}\eps^2}
    + \mathcal{O}(\eps^{-1}) ,
\end{align} \label{I431} \end{subequations}
\vspace{-5pt}
\begin{subequations} \begin{align}
 & I^{4-2\eps} \bigg(\usegraph{9}{delta332NPi}\bigg) [F_1] =
      \mathcal{O}(\eps^{-1}) , \\
 & I^{4-2\eps} \bigg(\usegraph{9}{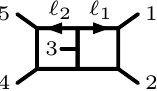}\bigg)
      [F_1\,(\ell_1\!\cdot\!(p_5-p_4))] =
      \mathcal{O}(\eps^{-1}) , \\
 & I^{4-2\eps} \bigg(\usegraph{9}{delta332NPl}\bigg)
      [F_1\,((\ell_1-\ell_2)\!\cdot\!p_3)] =
      \mathcal{O}(\eps^{-1}) ,
\end{align} \label{I332} \end{subequations}
\vspace{-5pt}
\begin{subequations} \begin{align}
 & I^{4-2\eps} \bigg(\!\!\!\:\usedelta{delta422NPi}\!\!\!\:\bigg) [F_1] =
    \frac{1}{(4\pi)^4} \frac{D_s-2}{3s_{12}s_{23}\eps^2} + \mathcal{O}(\eps^{-1}) , \\
 & I^{4-2\eps} \bigg(\!\!\!\:\usegraph{9}{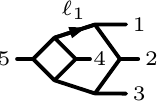}\!\!\!\:\bigg)
      [F_1 (\ell_1\!\cdot\!(p_5-p_4))] =
    \frac{1}{(4\pi)^4} \frac{(D_s-2)(s_{15}-s_{14}+s_{34}-s_{35})}
           {12 s_{12}s_{23}\eps^2} + \mathcal{O}(\eps^{-1}) ,
\end{align} \label{I422} \end{subequations}
\begin{align}
 & I^{4-2\eps} \bigg(\usegraph{13}{delta3315Li}\bigg) [F_1] =
    \frac{1}{(4\pi)^4} \frac{D_s-2}{6\eps^2}
      \left( \frac{1}{s_{12}} + \frac{1}{s_{45}} \right)
    + \mathcal{O}(\eps^{-1}) ,
\end{align}

\begin{align}
 & I^{4-2\eps} \bigg(\!\!\!\;\usegraph{16}{delta3225LNPi}\bigg) [F_1] =
    \frac{1}{(4\pi)^4} \frac{D_s-2}{6s_{12}\eps^2}+ \mathcal{O}(\eps^{-1}) ,
\end{align}

\begin{align}
 & I^{4-2\eps} \bigg(\usegraph{9}{delta331M1i}\bigg) [F_1] =
    \frac{1}{(4\pi)^4} \frac{D_s-2}{6s_{45}\eps^2} + \mathcal{O}(\eps^{-1}) ,
\end{align}

\begin{align}
 & I^{4-2\eps} \Delta\bigg(\usegraph{9}{delta232NPi}\!\bigg) [F_1] =
    \frac{1}{(4\pi)^4} \frac{D_s-2}{6s_{45}\eps^2} + \mathcal{O}(\eps^{-1}) ,
\end{align}

\begin{align}
  & I^{4-2\eps} \bigg(\!\!\!\:\usegraph{8.8}{delta322M1NPi}\bigg) [F_1] =
    \mathcal{O}(\eps^{-1}) .
\end{align}

Using these results, we have checked that \eqn{eq:IRsoft} does hold as expected for our amplitude~\eqref{A5point2loop2}.

\section{Conclusions}

In this paper we have explored the impact of tree-level amplitude relations
in multi-loop integrand computations. There were two major aspects to our work. Firstly, we exploited the Kleiss-Kuijf relations
to find a compact colour decomposition for the two-loop amplitude in terms of multi-peripheral colour factors in an analogous way
to the tree-level and one-loop decompositions of Del Duca, Dixon and Maltoni~\cite{DelDuca:1999ha,DelDuca:1999rs}.

Secondly, we applied the BCJ amplitude relations~\cite{Bern:2008qj}
to relate all non-planar generalized unitarity cuts
to the previously computed planar ones. This allowed us to easily generate
a compact representation the full colour two-loop, five-gluon, all-plus integrand building on previous planar work~\cite{Badger:2013gxa}.
The soft infrared poles of the full amplitude were checked
against the well-known universal pole structure.

We hope that the computational methods developed here will be of good use in the
necessary extension to more general helicity configurations and other $2\to3$
scattering processes at two loops. They highlight some advantages of relating
two-loop integrands to tree-level amplitudes via generalized unitarity cuts.
As well as avoiding the large intermediate steps that make Feynman diagram computations
at this loop order and multiplicity extremely computationally intensive, we are able to
build known on-shell symmetries and relations into the amplitude by construction.

Another interesting aspect of the all-plus amplitude is the continuing
connection to the previously known amplitudes in $\cN=4$ sYM. Though the
dimension shifting relation observed at one loop no longer holds, the integrands
of our full all-plus amplitude and the expressions of Carrasco and Johansson~\cite{Carrasco:2011mn}
are related by the same dimension shifting operator seen in the planar case. For example, we find that
\beal \!\!\!
   \Delta_{xyz;T}(1^+\!,2^+\!,3^+\!,4^+\!,5^+) =
      \frac{F_1(\mu_{11},\mu_{22},\mu_{12})}{\braket{12}^4}
      \Delta^{[\cN=4]}_{xyz;T}(1^-\!,2^-\!,3^+\!,4^+\!,5^+), \quad~ z \neq 0 .
\label{eq:2Ldimshift}
\eeal
The one-loop squared topologies
have the form
\beal \!\!\!\!
   \Delta_{xy0;T}(1^+\!,2^+\!,3^+\!,4^+\!,5^+) =
      \left( F_1(\mu_{11},\mu_{22},\mu_{12})
            -F_1(\mu_{11},\mu_{22},-\mu_{12}) \right)
      A(\{p_i\},\ell_1,\ell_2) & \\
    + F_3(\mu_{11},\mu_{22},\mu_{12}) B(\{p_i\},\ell_1,\ell_2) &, \!\!\!\!
\label{eq:butterflystructure}
\eeal
where $A$ and $B$ are some functions of the external kinematics and loop momenta.
The second term is proportional to $(D_s-2)^2$ and is a genuine contribution in QCD not related to $\cN=4$.
This additional numerator structure is enough
to make the off-shell BCJ symmetries non-trivial to satisfy,
even though the $\cN=4$ integrand has been cast in such a form.
It is an interesting question as to whether this would be possible
for the amplitudes presented here
and one that we intend to explore in the future.

\begin{acknowledgments}
  We would like to thank JJ Carrasco, Claude Duhr, Einan Gardi, and Henrik Johansson for useful discussions. Particular thanks go to Chris White for enlightening remarks on the soft singularities of loop integrals.
  SB is supported by an STFC Rutherford Fellowship ST/L004925/1. GM is supported by an STFC Studentship ST/K501980/1. AO is supported by the EU via a Marie Curie Actions grant FP7-PEOPLE-2013-CIG 631370.
  DOC is supported in part by the STFC consolidated grant ``Particle Physics at the Higgs Centre'', by the National Science Foundation under grant NSF PHY11-25915, and by the Marie Curie FP7 grant 631370.
\end{acknowledgments}

\bibliographystyle{JHEP}
\bibliography{references}

\end{document}